# A packet-based dual-rate PID control strategy for a slow-rate sensing Networked Control System.

A. Cuenca, J. Alcaina, J. Salt, V. Casanova, R. Pizá

Instituto Universitario de Automatica e Informatica Industrial,

Universitat Politecnica de Valencia, Spain

acuenca@isa.upv.es, joalac@upv.es, {julian, vcasanov, rpiza}@isa.upv.es

## Abstract

This paper introduces a packet-based dual-rate control strategy to face time-varying network-induced delays, packet dropouts and packet disorder in a Networked Control System. Slow-rate sensing enables to achieve energy saving by reducing network load. In addition, choosing a slower sensing period than the longest round-trip time delay can avoid packet disorder. On the other hand, a slow-rate sensing usually degrades control performance in a conventional control framework. Therefore, including dual-rate control techniques can be useful to maintain the desired performance, since the controller is able to generate a fast-rate control signal from a slow-rate sensing signal. A dual-rate PID controller is used, which can be split into two parts: a slow-rate PI controller is located at the remote side (with no permanent communication to the plant) and a fast-rate PD controller, at the local side (close to the plant, sensor, and inside the actuator, which can offer a low computation power). In addition, at the remote side, where a powerful computation device is located, a prediction stage is included in order to generate the packet of future, estimated slow-rate control actions to be sent to the local side. At this side, these actions are converted to fast-rate ones and used when a packet does not arrive due to the network-induced delay or due to occurring dropouts. The control proposal is able to reach the nominal (no-dropout, no-delay) performance despite the existence of time-varying delays and packet dropouts. Via real-time control for a Cartesian robot, results clearly reveal the superiority of the control approach compared to a previous authors' proposal, where the time-varying delays are faced by means of a gain scheduling control strategy.

## 1. - Introduction

One of the main goals of this contribution is to achieve energy saving in a low cost distributed sensor net. The sensors send data via a shared communication medium to a powerful remote server, that is, with high computation and information storage features. The interest of the work is to deal with a process control system by means of this kind of set-up. This is what is called a Networked Control System (NCS) [1,2,3]. Some interesting applications have been reported in this field [4,5,6,7] in the last few years. As it is well known, the energy consumption in sensor networks is usually due to sensing, processing and communication operations. Data transmission consumes most part of this amount of energy. Energy saving is a crucial issue, especially in battery-powered wireless sensors, having positive effects on their battery lives. Therefore, in order to save energy, a good option is to reduce the data flow frequency through the net. However, a quite low transmission frequency could imply failure to fulfil control specifications, or even process instability. Different control proposals have been introduced to reduce the communication rate preserving the control performance. One of them is the packet-based approach [8] which enables a sequence of signals to be sent over the network simultaneously. In the same way, techniques based on sending data when some thresholds are exceeded [9,10,11] –i.e. event-based control-, or just considering some specific data priority [12], have been used. Another possible option is the so called "multi-rate control", which is a control technique able to assume different rates for different control loop signals [13,14]. The present work combines multi-rate and packet-based control techniques for dealing with time-varying network-induced delays, packet dropouts, and packet disorder in an NCS.

The considered NCS locates the low cost sensor net, process and actuator (with a low computation power) in the local side, whereas the server (with high computation power and other capabilities) is situated in the remote side. Sensed values must travel through the main network from the local side to the remote side, and control signals, from the remote side to the local side. As it can be said, to reduce energy, a slow transmission frequency was assumed in both links. Therefore, slow-rate sensed values and slow-rate control actions travel through the NCS. Adopting a dual-rate control strategy, an *N* faster control updating can be assumed at the actuator device by converting the slow-rate control signal into a fast-rate one, which enables to achieve the desired control performance. This proposal requires a special controller, actually a non-conventional controller, which is able to compute control actions at period *T* from signals taken at period *NT* [14]. In this work, due to the wide knowledge of PID controllers in industrial and academic environments, a non-conventional PID control structure is taken into account. The controller is split into two parts: a slow-rate PI controller and a fast-rate PD controller. The integral action is applied at slow rate because it usually operates at this frequency zone, and hence, it is located at the remote (server) side. The derivative actions, which are associated with faster dynamics, are applied at fast rate, and hence, the PD controller is located at the local side (inside the actuator). The basic design procedure can be looked up in [14,15,16].

However, there are additional difficulties caused by the shared communication medium: network-induced delays and/or packet dropouts and/or packet disorder can appear depending on the network protocol used in a specific application. Regarding the delays, these are due to waiting-transmission-picking up times and are time-varying during the course of the application. As it is well-known, not compensating for the delays can imply a worsening of the control performance. This problem can be overcome by using, for example, gain-scheduling control strategies such as the one used in [17]. This proposal enables to obtain of a simple delay-dependent control law to retune the parameters of the fast-rate PD controller from the round-trip time delay. As the actuator is assumed to include low computation capabilities, it is able to measure the round-trip time delay and compute the fast-rate control signal. In this way, the nominal (no-delay) control performance can be closely maintained despite actuating in a non-uniform way due to the delays (that is, the last control action computed in the previous sensor period is held until new, current control actions are applied after the delay). Nevertheless, in the present work a packet-based control strategy is proposed, which enables to reach the nominal control performance not requiring any compensation for the delay. The packet received at the actuator includes future, estimated control information to compute the new control signal, which can be applied following a uniform actuation pattern (that is, actuating from the beginning of the current sensor period, in spite of the delay).

Concerning the packet disorder, once again, the fact of choosing a slow sensing period *NT* (concretely a slower sensing period than the longest round-trip time delay) is beneficial, since it can avoid this phenomenon. Obviously, the decision needs to perform some off-line experiences with usual operation conditions to detect some delay features. Sometimes, a statistical distribution of the network-induced delay is even found [18,19].

Finally, with regard to packet dropouts, if a network protocol like UDP is assumed, this phenomenon can occur [18,19]. In this work, a packet dropout can be derived either from an effective loss or from the expiration of a maximum waiting time. In addition, an upper-bound for consecutive packet dropouts *M* can be established from the off-line experiences.

In order to deal with up to *M* possible packet losses, a prediction stage is included at the remote side, which contains a state resetting procedure [20,21] in order to face even unstable processes (such as the position output for DC motors governing the axis of a Cartesian robot, used in this work). If a packet dropout occurs, a different solution is implemented for each network link. For the local-to-remote link, the process estimator provides the estimated output value in order to generate the current (estimated) PI control action. For the remote-to-local link, a packet-based strategy is adopted. The packet includes the current PI control action and *M* future ones to be used just in case the dropout occurs. As previously commented, whereas the fast-rate PD controller proposed in [17] holds the last PD action until new ones are calculated (which can be actual or estimated actions, depending on packet dropout occurring, or not), the control solution introduced in this work applies the new estimated control signal from the beginning of the sensor period *NT*. If a new PI action arrives with some delay, it produces a new PD action that is injected immediately, discarding the estimated one. If the new PI action is lost, the estimated control signal continues being applied.

The paper is structured in the following sections. In section 2, the problem scenario is formally introduced. In section 3, control techniques used in the remote and local sides are presented. Simulation results in section 4 illustrate the benefits of the proposed control strategy by comparison with the approach presented in [17]. Section 5 validates the results using a real physical process (a Cartesian robot). Finally, conclusions close this contribution.

## 2. - Problem description

The proposed NCS is depicted in Figure 1, where the network is placed between the remote and local sides, and can introduce time-varying delays, packet dropouts and packet disorder. The round-trip time delay for the packet sampled at the instant *kNT* (where *T* is the actuation period, $k \in \mathbb{N}$ is the iteration at period $NT$ -which is the sensor period-, and $N \in \mathbb{Z}^+$ is a parameter known as multiplicity in a dual-rate control framework [14]) is defined as

$$\tau_k = \tau_k^{lr} + \tau_k^{rl} + \tau_k^{c}, \qquad (2.1)$$

where $\tau_k^c$ can be considered as a negligible computation time delay, $\tau_k^{lr}$ is the local-to-remote network-induced delay, and $\tau_k^{rl}$, the remote-to-local one. To avoid packet disorder, $\tau_k \in [0, \tau_{\max}]$ must fulfill $\tau_{\max} < NT$. Since in this work, an IP network which uses UDP as the transport layer protocol is taken into account, the distribution of the round-trip time delay is a constant plus a Gamma distributed random variable, whose shape and scale parameters change with load and network segment [19]. Usually, this distribution is approximated as a generalized exponential distribution [22], whose probability density function can take this form

$$P[\tau_k] = \begin{cases} \dfrac{1}{\phi} e^{\dfrac{-(\tau_k - \eta)}{\phi}}, & \tau_k \geq \eta \\ 0 & , \tau_k < \eta \end{cases}, \qquad (2.2)$$

being the expected value of the delay $E[\tau_k] = \phi + \eta$, and its variance $V[\tau_k] = \phi^2$. A feasible choice of $\eta$ is the median of the delay. $\phi$ can be easily approximated from $\eta$, and an experimental value of $E[\tau_k]$ or the mean. Note that a common timer is supposed to be shared by the local devices in such a way that all of them are perfectly synchronized. Then, $\tau_k$ can be measured subtracting packet sending and receiving times, not requiring time-stamping techniques.

As well-known, when using the UDP transmission model, packet dropouts appear. This phenomenon is essentially random [18], and hence, it can be modeled as a Bernoulli distribution [1]. The variable $d_k^{lr}$ indicates the possible loss of the packet sent from the local side to the remote one at the instant $kNT$ (similarly, $d_k^{rl}$ is defined for the opposite network link). In this work, both variables are considered as a Bernoulli process with probability of dropout:

$$\begin{aligned} p^{lr} &= \Pr[d_k^{lr} = 0] \in [0,1) \\ p^{rl} &= \Pr[d_k^{rl} = 0] \in [0,1) \end{aligned}, \quad (2.3)$$

In some real scenarios, $p^{lr}, p^{rl}$ could be considered as the same value $p = p^{lr} = p^{rl}$. In the present study, $M$ is assumed as the upper bound of consecutive packet dropouts.

Next, the different devices included in Figure 1 for the NCS are presented:

- a process to be controlled: as it will be introduced in section 4 and 5, a Cartesian robot will be used.

- a sensor, working at period $NT$, to sample the process output $y_k^{NT}$. Sensing at this slow rate enables to achieve energy saving by reducing network load.

- a slow-rate PI controller, which generates a PI control action $u_{PI,k}^{NT}$ from the reference $r_k^{NT}$ and the sample $y_k^{NT}$, as long as it arrives to the remote side ($d_k^{lr} = 1$) after $\tau_k^{lr}$. Otherwise ($d_k^{lr} = 0$), a previously estimated PI control action $\hat{u}_{PI,k}^{NT}$ will be used. Note that to detect a packet dropout in this device, a maximum waiting time $\tau_{\max}^{lr}$ is considered. If $\tau_{\max}^{lr}$ expires and the packet does not arrive, it will be considered as a dropout. More information about the definition and operation mode of the slow-rate PI controller can be found in subsection 3.1.

- a prediction stage, which computes an array of $M$ estimated, future PI control actions $\left[ \hat{u}_{PI,k+1}^{NT}, \hat{u}_{PI,k+2}^{NT}, \ldots \hat{u}_{PI,k+M}^{NT} \right]$ from the array of the actual and future references $\left[ r_k^{NT}, r_{k+1}^{NT}, r_{k+2}^{NT}, \ldots r_{k+M}^{NT} \right]$, and the actual (or estimated) PI control action $u_{PI,k}^{NT}$ (or $\hat{u}_{PI,k}^{NT}$), the process output $y_k^{NT}$ (or $\hat{y}_k^{NT}$), and the state $x_k^{NT}$ (or $\hat{x}_k^{NT}$). For the sake of simplicity and brevity, both cases (actual and estimated) will be contained under the notation $\dot{u}_{PI,k}^{NT}, \dot{y}_k^{NT}, \dot{x}_k^{NT}$ on the sequel. The main goal of the prediction stage is facing packet dropouts for both network links. A more detailed scheme about this stage is depicted in Figure 2, where a prediction cascade structure is considered. For more information about how the prediction stage works, see subsection 3.4.

- a packet generator, which enables to consider a packet-based control strategy by creating the packet to be sent to the local side, containing $\left[ \dot{u}_{PI,k}^{NT}, \hat{u}_{PI,k+1}^{NT}, \hat{u}_{PI,k+2}^{NT}, \ldots \hat{u}_{PI,k+M}^{NT} \right]$

- an actuator which includes a rate converter and a fast-rate PD controller: the rate converter converts the slow-rate PI control signal to a fast-rate one in order to be used as an input to the fast-rate PD controller (more details in subsection 3.2). Then, the controller generates the PD control signal to be applied to the process by the actuator, which uses different actuation patterns according to $d_k^{rl}$:

  a) if dropout occurs ($d_k^{rl} = 0$): the actuator injects the control actions following a uniform pattern at time instants $\{0, T, 2T, \ldots, (N-1)T\}$ of the current sensor period $NT$. In this case, the PD control signal is an estimated one $\left[ \hat{u}_{PD,k}^{T}, \hat{u}_{PD,k+1}^{T}, \ldots, \hat{u}_{PD,k+N-1}^{T} \right]$, since, in order to be generated, an estimated PI control value is used from the previously stored packet.

  b) if no dropout occurs ($d_k^{rl} = 1$): the actuator injects the control actions following a non-uniform pattern. For simplicity, let us assume $\tau_k < T$. Then, the actuation time instants are $\{0, \tau_k, T, 2T, \ldots, (N-1)T\}$ inside the current sensor period $NT$. In this case, the PD control signal takes this form $\left[ \hat{u}_{PD,k}^{T}, \dot{u}_{PD,k}^{T}, \dot{u}_{PD,k+1}^{T}, \ldots, \dot{u}_{PD,k+N-1}^{T} \right]$, which means the injection of an estimated PD control action $\hat{u}_{PD,k}^{T}$ at the beginning of the current sensor period, and $N$ actual or estimated PD actions $\dot{u}_{PD,k+i}^{T}, i = [0..N-1]$ at the rest of the instants (from $\tau_k$).

  As previously, the notation $\dot{u}_{PD,k+i}^{T}$ is used to emphasize the possibility of having actual or estimated values. If both packets travelling through each network link arrive, the fast-rate PD controller can generate an actual control signal $u_{PD,k+i}^{T}$. But, if the packet travelling from the local to the remote sides is lost ($d_k^{lr}=0$), an estimated PI control action must be used to generate the current (and hence, estimated) PD control signal $\hat{u}_{PD,k+i}^{T}$.

The two main goals of the fast-rate PD controller are: 1) to achieve the required performance from a slow-rate sensing but acting at a fast rate (that is, a dual-rate control strategy [14]), 2) to compensate for the round-trip time delay, which can be measured at the local side.

Regarding the second goal, it is interesting to note that the proposed controller applies control actions from the time instant 0 of the current sensor period *NT* in spite of the delay (even when $d_k^{rl}=1$, where if a null estimation error were achieved, $\hat{u}_{PD,k}^{T}$ would be equal to $\dot{u}_{PD,k}^{T}$). Thus, no retuning for the controller according to the delay is required. This is an important improvement of the control approach proposed in this work compared to previous authors' proposals [15-17], where the last control action $\dot{u}_{PD,k+N-1}^{T,\tau_k}$ of the previous sensor period *NT* is held while no new one is applied due to the delay $\tau_k$. For this reason, in [15-17] the new *N* control actions $\left[\dot{u}_{PD,k}^{T,\tau_k},\dot{u}_{PD,k+1}^{T,\tau_k},\ldots,\dot{u}_{PD,k+N-1}^{T,\tau_k}\right]$ must be retuned according to the delay and adopting a gain-scheduling strategy. Figure 3 illustrates the difference between both approaches when no packet dropout occurs from remote to local sides ($d_k^{rl}=1$). In this figure, and henceforth, as the fast-rate PD controller proposed in this work does not depend on the round-trip time delay $\tau_k$, it is mentioned as delay-independent PD controller, in contrast to the ones presented in [15-17], which are referred to as delay-dependent PD controllers.

Figure 4 compares the other case, that is, when packet dropout occurs from remote to local sides ($d_k^{rl} = 0$). Originally, this case is not contemplated in [15-17], but including the predictor stage, the packet dropout phenomenon can also be treated. Then, the delay-dependent PD controller needs, firstly, to detect the dropout after expiring a supposed maximum waiting time $\tau_{max}$; secondly, to retune the controller according to $\tau_{max}$; and finally, to apply the control signal $\left[ \hat{u}_{PD,k}^{T,\tau_k}, \hat{u}_{PD,k+1}^{T,\tau_k}, \ldots, \hat{u}_{PD,k+N-1}^{T,\tau_k} \right]$ in a non-uniform way (considering $\tau_k = \tau_{max}$). Note, $\tau_{max}$ could fulfill $\tau_{max} \geq dT, d \in \mathbb{Z}^+$, then the first $d$ control actions would not be applied, and hence, the control performance could worsen. This is a possible problem solved by the new delay-independent PD controller, since, as previously commented in a) (if $d_k^{rl} = 0$), it is able to inject $N$ estimated control actions following a uniform pattern (and getting rid of the maximum delay).

More information about the fast-rate PD controller can be found in subsection 3.3, where the delay-independent approach is defined and compared to the delay-dependent one.

## 3. - Packet-based control strategy. Prediction stage.

In this section, the packet-based control strategy proposed in this work is formulated in subsections 3.1 (slow-rate PI controller), 3.2 (rate converter), and 3.3 (fast-rate PD controller). When defining the fast-rate PD controller (the so-called delay-independent PD controller), it will be concretely compared to that one presented in [17] (the so-called delay-dependent PD controller). Since packet dropouts could occur in both network links, each control stage must consider two cases: a) no packet dropout, b) packet dropout. At the end of the section (in subsection 3.4), the prediction stage is enunciated, considering the case of using both the delay-independent PD controller and the delay-dependent one.

First of all, let us define the transfer function of the continuous plant to be controlled as $G_p(s)$. By using the Z-transform at different periods plus a zero order hold device $H(s)$, different discrete-time versions for $G_p(s)$ can be calculated:

$$G^{NT}(z_N) \triangleq Z_{NT}\left[H_{NT}G_p(s)\right] = \frac{Y^{NT}(z_N)}{U^{NT}(z_N)}; \quad G^T(z) \triangleq Z_T\left[H_T G_p(s)\right] = \frac{Y^T(z)}{U^T(z)}$$

$$G^t(\bar{z}) \triangleq Z_t\left[H_t G_p(s)\right] = \frac{Y^t(\bar{z})}{U^t(\bar{z})}, \quad t < T: \ t \cdot L = T, \ L \in \mathbb{Z}^+,$$

(3.1)

In addition, the consequent state-space representations for each case can be enunciated as

$$\begin{cases} x_{k+1}^{NT} = A^{NT} x_k^{NT} + B^{NT} u_k^{NT} \\ y_k^{NT} = C^{NT} x_k^{NT} \end{cases}; \begin{cases} x_{k+1}^T = A^T x_k^T + B^T u_k^T \\ y_k^T = C^T x_k^T \end{cases}; \begin{cases} x_{k+1}^t = A^t x_k^t + B^t u_k^t \\ y_k^t = C^t x_k^t \end{cases}, \quad (3.2a)$$

Secondly, let us consider a continuous PID which is designed according to classical methods in order to achieve certain specifications for the process to be controlled. This is the configuration considered for the continuous PID controller:

$$G_{PID}(s) = K_p\left(1 + T_d s + \frac{1}{T_i s}\right), \quad (3.2b)$$

## 3.1. Slow-rate PI controller

Since packet dropouts could occur through the local-to-remote link, the following two cases must be considered:

a) No dropout ($d_k^{lr} = 1$): The PI controller working at period *NT* is enunciated as

$$G_{PI}^{NT}(z_N) = K_{PI} \frac{z_N - \left(1 - \frac{NT}{T_i}\right)}{z_N - 1} = \frac{U_{PI}^{NT}(z_N)}{E^{NT}(z_N)}, \quad (3.3)$$

being $U_{PI}^{NT}(z_N)$ the PI control signal, $E^{NT}(z_N)$ the error signal, and $K_{PI}, T_i$ the gains of the PI controller (usually, $K_{PI} = 1$). The PI control signal is obtained as

$$U_{PI}^{NT}(z_N) = G_{PI}^{NT}(z_N) E^{NT}(z_N) = G_{PI}^{NT}(z_N)\left(R^{NT}(z_N) - Y^{NT}(z_N)\right), \quad (3.4)$$

and, from (3.3), the difference equation for the PI controller (with $K_{PI} = 1$) will be

$$u_{PI,k}^{NT} = u_{PI,k-1}^{NT} + e_{PI,k}^{NT} - \left(1 - \frac{NT}{T_i}\right) e_{PI,k-1}^{NT} = u_{PI,k-1}^{NT} + \left(r_k^{NT} - y_k^{NT}\right) - \left(1 - \frac{NT}{T_i}\right)\left(r_{k-1}^{NT} - y_{k-1}^{NT}\right), \quad (3.5)$$

b) Dropout ($d_k^{lr} = 0$): In this case, instead of using the actual PI control signal in (3.4), the estimated one $\hat{U}_{PI}^{NT}(z_N)$ must be used. This signal is previously generated at the prediction stage according to subsection 3.4.

*3.2. Rate converter*

As it is well-known [14], a rate converter $[H_{NT}]^T$ between slow (remote) and fast (local) controllers is required. Its goal is to convert the slow-rate PI control signal to a fast-rate one in order to be used as an input to the fast-rate PD controller. For practical purposes, when the reference to be followed by the plant is a step, the rate converter acts as a zero order hold. This operation can be carried out either at the remote side (sending the converted signal to the local side) or directly at the local side (this is the option used in this work). Two cases are considered depending on packet dropout occurring, or not, in the remote-to-local network link:

a) No dropout ($d_k^{rl} = 1$): The rate converter considers the actual slow-rate PI control signal $U_{PI}^{NT}(z_N)$ to obtain the held fast-rate one $\bar{U}_{PI}^T(z)$:

$$[H_{NT}]^T = \frac{\bar{U}_{PI}^T(z)}{\left[U_{PI}^{NT}\right]^T(z)} = \frac{1 - z^{-N}}{1 - z^{-1}} \to \bar{U}_{PI}(z) = [H_{NT}]^T \left[U_{PI}^{NT}\right]^T(z), \quad (3.6a)$$

Note that $U_{PI}^{NT}(z_N)$ is required to be used in an expanded way $\left[U_{PI}^{NT}\right]^T(z)$, that is,

$$\left[U_{PI}^{NT}\right]^T(z) \triangleq \tilde{U}_{PI}^T(z) \triangleq \sum_{k=0}^{\infty} \tilde{u}_{PI,k}^T z^{-k} : \begin{cases} \tilde{u}_{PI,k}^T = u_{PI,k}^T, \forall k = \lambda N \\ \tilde{u}_{PI,k}^T = 0, \quad \forall k \neq \lambda N \end{cases}, \lambda \in Z^+, \quad (3.6b)$$

More information in [14].

b) Dropout ($d_k^{rl} = 0$): Now, the rate converter considers the estimated PI control signal $\hat{U}_{PI}^{NT}(z_N)$:

$$\hat{\bar{U}}_{PI}(z) = [H_{NT}]^T [\hat{U}_{PI}^{NT}]^T (z), \tag{3.7}$$

As used in section 2, for the sake of simplicity and brevity, both cases ((3.6a) and (3.7)) will be contained under the notation $\dot{\bar{U}}_{PI}^{T}(z)$ from now on.

*3.3. Fast-rate PD controller*

Once again, two cases are treated (no dropout versus dropout in the remote-to-local network link) but now, for each case, the two different control approaches will be presented (delay-dependent controller versus delay-independent controller).

a) No dropout ($d_k^{rl} = 1$)

1. Delay-dependent controller: After (2.1), the round-trip time delay was defined to fulfill $\tau_k \in [0, \tau_{max}]$. As commented in section 2, let us suppose $\tau_{max} < T$ (for example, $\tau_{max} = T - t$) in order to inject $N$ control actions for each sensor period $NT$ (remember Figure 3). The PD controller, working at period $T$, is enunciated as

$$G_{PD}^{T,\tau_k}(z) = K_{PD}^{\tau_k} \frac{z\left(1 + \frac{T_d^{\tau_k}}{T}\right) - \frac{T_d^{\tau_k}}{T}}{z} = \frac{\dot{U}_{PD}^{T,\tau_k}(z)}{\dot{\bar{U}}_{PI}^{T}(z)} \rightarrow \dot{U}_{PD}^{T,\tau_k}(z) = G_{PD}^{T,\tau_k}(z)\dot{\bar{U}}_{PI}^{T}(z), \tag{3.8}$$

where $K_{PD}^{\tau_k}, T_d^{\tau_k}$ are the gains of the PD controller (usually, $K_{PD}^{\tau_k} = K_p^{\tau_k}$), which are retuned according to $\tau_k$ via the gain-scheduling algorithm presented in [17]. Note that the notation $\dot{U}_{PD}^{T,\tau_k}(z)$ represents the PD control signal obtained either from the actual PI control signal $\bar{U}_{PI}^{T}(z)$ (in (3.6a)) or from its estimation $\hat{\bar{U}}_{PI}^{T}(z)$ (in (3.7)). At the current sensor period, in addition to the last PD control action of the previous period (which remains held), new $N$ PD control actions are applied to the plant after $\tau_k$ (remember Figure 3). These $N$ actions are obtained after iterating the difference equation deduced from (3.8) $N$ times, that is

$$\dot{u}_{PD,k}^{T,\tau_k} = K_{PD}^{\tau_k}\left(1+\frac{T_d^{\tau_k}}{T}\right)\dot{\bar{u}}_{PI,k}^T - K_{PD}^{\tau_k}\left(\frac{T_d^{\tau_k}}{T}\right)\dot{\bar{u}}_{PI,k-1}^T, \tag{3.9}$$

As commented in section 2, due to the delay, these actions will be applied following a non-uniform pattern. Then, a basic period $t$ is required to adapt the non-uniformity to the delay in such a way that the actuation pattern inside the sensor period $NT$ will take this form (where $l=0..LN-1$):

$$\begin{cases} \dot{u}_{PD,k-1}^{T,\tau_k}, & lt = 0..\tau_k \\ \dot{u}_{PD,k}^{T,\tau_k}, & lt = \tau_k..T \\ \dot{u}_{PD,k+1}^{T,\tau_k}, & lt = T..2T \\ \vdots \\ \dot{u}_{PD,k+N-1}^{T,\tau_k}, & lt = (N-1)T..NT \end{cases}, \tag{3.10}$$

2. Delay-independent controller: In this case, the gains of the PD controller $K_{PD}, T_d$ (usually $K_{PD} = K_p$) do not depend on the delay $\tau_k$. Then, the controller is defined as

$$G_{PD}^T(z) = K_{PD}\frac{z\left(1+\frac{T_d}{T}\right)-\frac{T_d}{T}}{z} = \frac{\dot{U}_{PD}^T(z)}{\dot{\bar{U}}_{PI}^T(z)} \rightarrow \dot{U}_{PD}^T(z) = G_{PD}^T(z)\dot{\bar{U}}_{PI}^T(z), \tag{3.11}$$

and its difference equation

$$\dot{u}_{PD,k}^T = K_{PD}\left(1+\frac{T_d}{T}\right)\dot{\bar{u}}_{PI,k}^T - K_{PD}\left(\frac{T_d}{T}\right)\dot{\bar{u}}_{PI,k-1}^T, \tag{3.12}$$

From (3.12), the $N$ PD control actions are generated and applied after $\tau_k$ (when $\dot{\bar{U}}_{PI}^T(z)$ is available). As depicted in Figure 3, unlike (3.10), where the last action of the previous sensor period is held before applying the $N$ actions, now the action injected at the beginning of the sensor period $NT$ is obtained according to (3.16), that is, from the estimated PD control signal $\hat{U}_{PD}^T(z)$. Therefore, the non-uniform actuation pattern inside the sensor period $NT$ will be (where $t$ is the basic period, and $l=0..LN-1$):

$$\begin{cases} \hat{u}^T_{PD,k}, & lt = 0..\tau_k \\ \hat{u}^T_{PD,k}, & lt = \tau_k..T \\ \hat{u}^T_{PD,k+1}, & lt = T..2T \\ \vdots \\ \hat{u}^T_{PD,k+N-1}, & lt = (N-1)T..NT \end{cases}, \tag{3.13}$$

b) Dropout ($d^{rl}_k = 0$): The estimated PI control signal $\hat{U}^T_{PI}(z)$ (to be defined in the last step in subsection 3.4) is now required. This control signal is available at the local side, since it was calculated at the remote side in a previous iteration and sent to the local side in a previous successful communication.

1. Delay-dependent controller: the time $\tau_{max} = T - t$ is considered as the maximum waiting time established to detect a packet dropout (remember Figure 4). In this way, $N$ PD control actions (in addition to the held action from the previous sensor period) are guaranteed to be applied in the current sensor period (as in (3.10)). In this case, with certainty, the PD control signal is an estimated one, since it depends on the estimated PI control signal. Then, with $\tau_k = \tau_{max}$

$$\hat{U}^{T,\tau_k}_{PD}(z) = G^{T,\tau_k}_{PD}(z)\hat{\bar{U}}^T_{PI}(z), \tag{3.14}$$

From (3.9), but considering estimated signals, the set of $N$ control actions can be computed and applied according to the next non-uniform actuation pattern inside the sensor period $NT$ (where $t$ is the basic period, and $l=0..LN-1$):

$$\begin{cases} \hat{u}^{T,\tau_k}_{PD,k-1}, & lt = 0..\tau_k \\ \hat{u}^{T,\tau_k}_{PD,k}, & lt = \tau_k..T \\ \hat{u}^{T,\tau_k}_{PD,k+1}, & lt = T..2T \\ \vdots \\ \hat{u}^{T,\tau_k}_{PD,k+N-1}, & lt = (N-1)T..NT \end{cases}, \tag{3.15}$$

2. Delay-independent controller: As depicted in Figure 4, in this case, in spite of the delay, the first of the $N$ control actions is applied at the beginning of the current sensor period. The estimated PD control signal takes this form

$$\hat{U}_{PD}^{T}(z) = G_{PD}^{T}(z)\hat{\bar{U}}_{PI}^{T}(z),  \qquad (3.16)$$

From (3.12), but considering estimated signals, the set of *N* control actions are computed and applied according to the next uniform actuation pattern inside the sensor period *NT*:

$$\begin{cases} \hat{u}_{PD,k}^{T}, & kT = 0..T \\ \hat{u}_{PD,k+1}^{T}, & kT = T..2T \\ \vdots \\ \hat{u}_{PD,k+N-1}^{T}, & kT = (N-1)T..NT \end{cases}, \qquad (3.17)$$

*3.4. Prediction stage*

The prediction algorithm is executed *M* times (*M* was defined in section 2 as the upper bound of consecutive packet dropouts) following a cascade structure (remember Figure 2) in order to generate the packet which includes the future, estimated *M* PI control actions $\left[\hat{u}_{PI,k+1}^{NT}, \hat{u}_{PI,k+2}^{NT}, \ldots \hat{u}_{PI,k+M}^{NT}\right]$. This packet is computed for every sensor period at the remote side, and it is sent to the local side in order to be stored, and used if subsequent dropouts occur through the remote-to-local communication. Considering a for-loop where *i*=1..*M*, the statements of the prediction algorithm included in the loop are based on the next steps:

1. Resetting of the initial state: If the current state sensed at period *NT*, $x_k^{NT}$, is available at the remote side (that is, no dropout occurs when being sent via the local-to-remote network link), a resetting of the initial condition for the state at period *t* and *T* can be carried out. This operation can be executed when *i*=1, and it is required in order to deal with unstable plants or marginally stable plants [20, 21] such as the one used in this work (i.e. the position output for DC motors which govern the axis of a Cartesian robot). For the rest of iterations of the algorithm (*i*=2..*M*), or if the current state was dropped (for *i*=1), the updating is computed from the estimated state $\hat{x}_{k+i-1}^{NT}$ (to be defined in step 3). As in section 2, to contemplate every situation, let us define a generic (actual or estimated) state $\dot{x}_k^{NT}$. Therefore, the resetting carried out in each iteration is

$$\begin{cases} i=1: & \hat{x}_k^T \leftarrow \dot{x}_k^{NT}; \quad \hat{x}_k^t \leftarrow \dot{x}_k^{NT} \\ i>1: & \hat{x}_{k+(i-1)N}^T \leftarrow \hat{x}_{k+i-1}^{NT}; \quad \hat{x}_{k+(i-1)LN}^t \leftarrow \hat{x}_{k+i-1}^{NT} \end{cases}, \quad (3.18)$$

2. Estimation of the *N* PD control actions either from the estimated PI control signal $\hat{\bar{U}}_{PI}^T(z)$ (it can occur for $i \geq 1$) or from the actual one $\bar{U}_{PI}^T(z)$ (it can only occur for *i*=1). Both cases assume the previous rate conversion ((3.7) or (3.6a), respectively). The estimation of the PD control signal depends on which of the fast-rate PD controllers is considered:

a. Delay-dependent controller: Similarly to (3.9), the estimated control signal is computed by iterating the next difference equation for *j*=0..*N*-1. Each iteration *i* for the prediction algorithm is calculated as follows

$$\begin{cases} i=1: & \hat{u}_{PD,k+j}^{T,\tau_k} = K_{PD}^{\tau_k}\left(1+\dfrac{T_d^{\tau_k}}{T}\right)\dot{\bar{u}}_{PI,k+j}^T - K_{PD}^{\tau_k}\left(\dfrac{T_d^{\tau_k}}{T}\right)\dot{\bar{u}}_{PI,k-1+j}^T, \quad \tau_k = \tau_m \\ i>1: & \hat{u}_{PD,k+j+(i-1)N}^{T,\tau_k} = K_{PD}^{\tau_k}\left(1+\dfrac{T_d^{\tau_k}}{T}\right)\hat{\bar{u}}_{PI,k+j+(i-1)N}^T - K_{PD}^{\tau_k}\left(\dfrac{T_d^{\tau_k}}{T}\right)\hat{\bar{u}}_{PI,k-1+j+(i-1)N}^T, \quad \tau_k = \tau_{\max} \end{cases}, \quad (3.19)$$

where now, the equation for the first iteration of the prediction algorithm (*i*=1) is calculated supposing a successful remote-to-local communication (the packet which includes the estimated PI control actions will arrive to the local side) but unknowing the consequent remote-to-local delay (this information is not available at this moment) and hence, the round-trip time delay. This is the reason of adopting the statistical mode of the delay distribution $\tau_m$ as the delay to be considered at the first iteration. The mode can be obtained via a previous statistical analysis about the delay nature. For the rest of iterations (when *i*>1), the considered delay to generate the PD control actions is the maximum waiting time established to detect a dropout, that is, $\tau_k = \tau_{\max}$, since the next *M*-1 packets are assumed to be dropped.

b. Delay-independent controller: Similarly to (3.12), the estimated control signal is computed by iterating the next difference equation for $j=0..N-1$. Each iteration $i$ for the prediction algorithm is calculated as follows

$$\begin{cases} i=1: \ \hat{u}^T_{PD,k+j} = K_{PD}\left(1+\dfrac{T_d}{T}\right)\dot{\hat{u}}^T_{PI,k+j} - K_{PD}\left(\dfrac{T_d}{T}\right)\dot{\hat{u}}^T_{PI,k-1+j} \\ i>1: \ \hat{u}^T_{PD,k+j+(i-1)N} = K_{PD}\left(1+\dfrac{T_d}{T}\right)\dot{\hat{u}}^T_{PI,k+j+(i-1)N} - K_{PD}\left(\dfrac{T_d}{T}\right)\dot{\hat{u}}^T_{PI,k-1+j+(i-1)N} \end{cases}, \quad (3.20)$$

3. Estimation of the next state and output at period $NT$ from the estimated PD control actions. Once again, two cases can be considered depending on the fast-rate PD controller used in the previous step:

a. Delay-dependent controller: As in (3.10), (3.13) and (3.15), the basic period $t$ is used. In this case, for each iteration of the prediction algorithm $i=1..M$, the next state-space representation is calculated by iterating for $l=0..LN-1$

$$\begin{cases} \hat{x}^t_{k+1+l+(i-1)LN} = A^t \hat{x}^t_{k+l+(i-1)LN} + B^t \left\langle \hat{u}^{T,\tau_k}_{PD,k+l+(i-1)LN} \right\rangle^t \\ \hat{y}^t_{k+1+l+(i-1)LN} = C^t \hat{x}^t_{k+1+l+(i-1)LN} \end{cases}, \quad (3.21)$$

where, for simplicity and brevity, let us represent by means of $\left\langle \hat{u}^{T,\tau_k}_{PD,k+l+(i-1)LN} \right\rangle^t$ the sequence of the $N+1$ PD control actions included in each sensor period by holding them at period $t$, that is, applying them via a non-uniform actuation pattern such as the one used in (3.15) (in [15], a more accurate representation for (3.21) can be found). As a result of iterating (3.21) for all $i$, the $M$ estimated states and outputs at period $NT$, $\hat{x}^{NT}_{k+i}, \hat{y}^{NT}_{k+i}$, are obtained.

b. Delay-independent controller: Now, as in (3.17), a uniform pattern is used. Then, for each iteration of the prediction algorithm $i=1..M$, the next state-space representation at period $T$ is computed for $j=0..N-1$:

$$\begin{cases} \hat{x}^T_{k+1+j+(i-1)N} = A^T \hat{x}^T_{k+j+(i-1)N} + B^T \hat{u}^T_{PD,k+j+(i-1)N} \\ \hat{y}^T_{k+1+j+(i-1)N} = C^T \hat{x}^T_{k+1+j+(i-1)N} \end{cases}, \quad (3.22)$$

As a result of iterating (3.22) for all *i*, the *M* estimated states and outputs at period *NT*, $\hat{x}_{k+i}^{NT}, \hat{y}_{k+i}^{NT}$, are calculated.

4. Estimation of the PI control signal $\hat{U}_{PI}^{NT}(z_N)$ from the estimated output signal $\hat{Y}^{NT}(z_N)$. Note that, particularly for the first iteration of the prediction algorithm (*i*=1), the actual output $y_k^{NT}$ can be used if it is available at the remote side, that is, if no dropout occurs when being sent through the local-to-remote network link ($d_k^{lr} = 1$). Then, the actual PI control action $u_{PI,k}^{NT}$, which is generated by the output $y_k^{NT}$ (remember (3.5)), can also be used. In this way, similarly to step 1, a resetting of the initial condition for the PI controller ($u_{PI,k}^{NT}$) is carried out in order to compute the next estimated PI control action $\hat{u}_{PI,k+1}^{NT}$. This operation is useful due to the marginally stable open-loop nature of the PI controller [20, 21]. As usual, in order to contemplate every situation in the prediction algorithm, let us define a generic (actual or estimated) output $\dot{y}_k^{NT}$, and a generic (actual or estimated) control action $\dot{u}_{PI,k}^{NT}$. Therefore, similarly to (3.5), the iterations *i* of the prediction algorithm take the form

$$\begin{cases} i = 1: \quad \hat{u}_{PI,k+1}^{NT} = \dot{u}_{PI,k}^{NT} + \left(r_{k+1}^{NT} - \hat{y}_{k+1}^{NT}\right) - \left(1 - \frac{NT}{T_i}\right)\left(r_k^{NT} - \dot{y}_k^{NT}\right) \\ i > 1: \quad \hat{u}_{PI,k+i}^{NT} = \hat{u}_{PI,k+i-1}^{NT} + \left(r_{k+i}^{NT} - \hat{y}_{k+i}^{NT}\right) - \left(1 - \frac{NT}{T_i}\right)\left(r_{k+i-1}^{NT} - \hat{y}_{k+i-1}^{NT}\right) \end{cases}, \quad (3.23)$$

## 4. Simulation results

### *4.1. Simulation data*

In this section, a particular case for the proposed NCS is presented. The different control solutions are compared by simulation, which is based on a model of a plant available in the laboratory in order to then validate the results experimentally (in section 5).

The process selected to be controlled is a Cartesian robot manufactured by Inteco, specifically, the 3D CRANE module (see in Figure 5). The rail measures of this plant for each axis are: X=0.050m, Y=0.040m, Z=0.050m.

Focusing on the X axis, it is identified, obtaining the next model

$$G_p(s) = \frac{6.3}{s(s+17.7)} \text{ m/c.a.u.}, \quad (4.1)$$

where c.a.u. means control action units, which are generated by a PWM signal normalized in the range [0,1].

The system also presents two non-linear behaviors to be taken into account in real-time implementation: saturation limits of control actions in ±1, and dead zone values of ±0.06. Both of them are identified experimentally and measured in normalized c.a.u.

In [17], a conventional PID controller such as in (3.2b) is used with $K_p = 12$, $T_d = 0.01$ and $T_i = 3.5$, which reaches certain specifications (null steady-state error, settling time around 4s, and overshot around 5%). Digitally implementing this controller at a period higher than 0.1s, the aforementioned specifications cannot be assured. It is assumed that the sensor's nature or the network load do not allow a sampling period below 0.2s. Therefore, a sample time of NT=0.2s is used, and a dual-rate controller with N=2 is implemented using (3.3)-(3.17) (assuming $K_{PD} = K_p$ and $K_{PI} = 1$) in order to reach the specifications. The following gain-scheduling law for the dual-rate delay-dependent controller is used to deal with the round-trip time delay $\tau_k$:

$$\begin{aligned} K_{PD}^{\tau_k} &= -50\tau_k + K_{PD} \\ T_d^{\tau_k} &= 0.5\tau_k + T_d \end{aligned}, \quad (4.2)$$

For the present study, as enunciated in (2.2), a generalized exponential distribution is considered. The histogram used in this case is shown in Figure 6, where $\tau_k$ takes values in the range $\Phi = [0.04, 0.08]$. In addition, as presented in (2.3), packet dropouts are modeled as a Bernoulli distribution, being in this case $p = p^{lr} = p^{rl} = 0.3$ and M=3. Figures 7 and 8 show the results obtained for each control solution, where filtered step references are used in order to avoid the saturation of the control signal. Note that the sequence of packet dropouts is represented in the time axis in such a way that a blue point indicates a packet dropout in the time instant where it is plotted. If the point increases its value in the vertical axis, then consecutive dropouts are occurring in this instant. Both Figures 7 and 8 show the desired, nominal (no-delay, no-dropout) output. When packet dropouts appear in the NCS, if the delay-dependent control solution does not include the prediction stage, behavior deterioration is observed (Figure 7). Then, including the prediction stage, the control performance can be restored, but not accurately reached. This fact is a result of using a linear retune law to compensate for the delays, which works better with short delays than with longer ones (more information in [15,17]). Nevertheless, the delay-independent control solution (including the prediction stage) is able to achieve the desired control properties (Figure 8). Later, in section 4.2, all of these conclusions will be quantified.

Note that model-based control solutions are considered in this work. Therefore, both the controller design and the prediction computation depend on how precise the model represents the plant behavior. Assuming the existence of certain uncertainty between plant and model, the robustness of a model-based control proposal can be checked. The study can consider a modification in the characteristic parameters of the plant (say, the static gain $K$ and the time constant $\tau$) with regard to the previous ones used in the controller design and prediction stage. In this case, let us consider a percentage $q\%$ of decrement in $K$ ($q\%\Delta K$) and a percentage $r\%$ of decrement in $\tau$ ($r\%\Delta\tau$). Focusing on the delay-independent control solution (with prediction stage), Figure 9 shows two outputs obtained when the plant and the model differ certain $q\%\Delta K$ and $r\%\Delta\tau$. As expected, the higher the percentage of uncertainty is considered, the worse the behavior becomes (with regard to the nominal one). However, the control solution seems to be robust since, despite considering significant uncertainties (up to 30% in $K$ and up to 12% in $\tau$), the worsening seems not to be excessive (for example, the overshoot is increased around 6%, and the settling time around 60%). In the next subsection, in order to quantify this study in more detail, some cost indexes will be used.

*4.2. Data anylisis via cost indexes*

In this subsection four different cost indexes will be used. Firstly, in order to better quantify the benefits of the delay-independent control solution, the cost indexes $J_1$ and $J_2$ are utilized. The first one, $J_1$, is based on the Integral of Absolute Error (IAE), and the second one, $J_2$, on the overshoot value. $J_1$ and $J_2$ take the worst behavior in Figure 7 (which was obtained by the delay-dependent controller with no prediction stage) as the reference value to be compared to the rest of behaviors in order to compute the consequent improvements. Secondly, in order to quantify the robustness of the delay-independent control solution, the cost indexes $J_3$ and $J_4$ are used, which are also based on the IAE and on the overshoot, respectively. These indexes take the worst response in Figure 9 as the reference value to be compared to the rest of behaviors.

In order to define $J_1$ let us consider the array $Y$, which includes the different sequence of outputs to be analyzed for the dual-rate control system, that is, $Y = [Y_{Nom}, Y_{DD-NP}, Y_{DD-P}, Y_{DI-P}]$, being $Y_{Nom}$ the output for the nominal (no-delay, no-dropout) case, $Y_{DD-NP}$ the output for the delay-dependent controller with no prediction stage, $Y_{DD-P}$ the output for the delay-dependent controller with prediction stage, and $Y_{DI-P}$ the output for the delay-independent controller with prediction stage.

From $Y$ the following accumulated (integrated) error $E_Y$ in a range of time instants $\Gamma$ can be computed

$$E_Y(i) = \sum_{\Gamma} |Y(i) - Y_{Nom}|, \quad i = 1..4, \quad (4.3)$$

Then, the $J_1$ cost index takes this form

$$J_1(i) = 100 - \frac{E_Y(i)}{E_Y(2)} 100 \ (\%), \quad i = 1..4, \tag{4.4}$$

being $E_Y(2)$ the worst expected accumulated error, that is, the error reached by the delay-dependent controller with no prediction stage. Therefore, the rest of the errors are measured by $J_1$ as an improvement (in %) with regard to $E_Y(2)$.

In order to define $J_2$, from $Y$ the following overshoot $O_Y$ in a range of time instants $\Gamma$ can be calculated (considering positive -max- or negative -min- filtered step references)

$$O_Y(i) = \max\left(\left|\max_\Gamma Y(i) - \max_\Gamma Y_{Nom}\right|, \left|\min_\Gamma Y(i) - \min_\Gamma Y_{Nom}\right|\right), \quad i = 1..4, \tag{4.5}$$

Then, the $J_2$ index is

$$J_2(i) = 100 - \frac{O_Y(i)}{O_Y(2)} 100 \ (\%), \quad i = 1..4, \tag{4.6}$$

being $O_Y(2)$ the worst expected overshoot, that is, the overshoot reached by the delay-dependent controller with no prediction stage. Similarly to $J_1$, the rest of the overshoots are measured by $J_2$ as an improvement (in %) with regard to $O_Y(2)$.

Table 1 summarizes the cost indexes obtained for each output. As expected, considering packet dropouts, both packet-based control strategies which include prediction stage improve the control performance obtained by the no-prediction control strategy. Although the delay-dependent control solution with prediction stage significantly improves $J_2$, an elevated value for $J_1$ still appears. Nevertheless, the delay-independent control approach (with prediction stage) is able to accurately achieve the same control properties as the nominal dual-rate control solution.

In order to define $J_3$ let us consider the matrix $W$, which includes the different sequence of outputs to be analyzed for the delay-independent control system. These outputs are obtained as a result of varying $q\%\Delta K$ and $r\%\Delta\tau$. In this study, $q$ takes the values $q=0$, 20, 30, and $r$, the values $r=0$, 8, 12. Therefore, 9 different responses in $W$ are considered, being one of them the nominal one $Y_{Nom}$ (when $q=r=0$). As expected, the worst behavior will be obtained for $q=30$ and $r=12$, since it represents the highest divergence between model and plant in this study. Let us assume this behavior as the worst permissible one.

From $W$ the following accumulated (integrated) error $E_W$ in a range of time instants $\Gamma$ can be computed

$$E_W(i_r, i_q) = \sum_{\Gamma} |W(i_r, i_q) - Y_{Nom}|, \quad i_r, i_q = 1..3, \tag{4.7}$$

Then, the $J_3$ cost index is

$$J_3(i_r, i_q) = 100 - \frac{E_W(i_r, i_q)}{E_W(3,3)} 100 \, (\%), \quad i_r, i_q = 1..3, \tag{4.8}$$

being $E_W(3,3)$ the worst permissible accumulated error, that is, the error reached when $r=12$ and $q=30$. Therefore, the rest of the errors are measured by $J_3$ as an improvement (in %) with regard to $E_W(3,3)$.

In order to define $J_4$, from $W$ the following overshoot $O_W$ in a range of time instants $\Gamma$ can be calculated (considering positive -max- or negative -min- filtered step references)

$$O_W(i_r, i_q) = \max\left(\left|\max_{\Gamma} W(i_r, i_q) - \max_{\Gamma} Y_{Nom}\right|, \left|\min_{\Gamma} W(i_r, i_q) - \min_{\Gamma} Y_{Nom}\right|\right), \quad i_r, i_q = 1..3, \tag{4.9}$$

And then, the $J_4$ index is

$$J_4(i_r, i_q) = 100 - \frac{O_W(i_r, i_q)}{O_w(3,3)} 100 \, (\%), \quad i_r, i_q = 1..3, \tag{4.10}$$

being $O_W(3,3)$ the worst permissible overshoot, that is, the overshoot reached when $r=12$ and $q=30$. Similarly to $J_3$, the rest of the overshoots are measured by $J_4$ as an improvement (in %) with regard to $O_W(3,3)$.

Tables 2 and 3 summarize the cost indexes obtained for each output. As expected, the lower the percentage of divergence is considered, the higher $J_3$ and $J_4$ become, that is, a closer behavior to the nominal one is obtained.

# 5. - Experimental results

To validate the simulation results obtained in section 4, a laboratory test-bed platform is set up, which includes the CRANE module previously presented, two computers and an Ethernet cable. One computer is directly connected to the plant and composes the local part of the control system. The aims of this computer are: firstly, to be in charge of the measures on the plant at $NT=0.2$s and their communication; secondly, to be responsible for the reception of the slow-rate control signal from the PI controller, the calculation of the fast-rate PD control actions (at $T=0.1$s) and their injection over the plant. When convenient, it is also in charge of the reception and selection of signals predicted to be applied to the plant when loss of communication occurs.

The second computer performs the remote part of the controller, receiving the outputs of the plant, calculating the slow-rate PI controller, and sending back these actions to the local system. When required, this part is also in charge of the calculation of future, slow, predicted control actions, which will be sent together with the slow-rate control signal.

These computers are connected by a UDP network through an Ethernet cable that performs the local-to-remote and remote-to-local links. In order to obtain similar conditions to those considered in simulation, packet delays and packet dropouts are modified by software.

Figures 10 and 11 show the outputs obtained in the experiment, which clearly reveals the same trend observed in Figures 7 and 8 respectively. To better validate the results, Table 4 details the cost indexes $J_1$ and $J_2$ computed for the experiment, where lower improvements than in simulation are achieved due to practical issues (possible divergences between model and plant, dead zone, and so on).

The Cartesian robot is able to track 2D and 3D trajectories. For the sake of clarity let us show a 2D trajectory based on the well-known Lissajous curves (see, for example, in [23]). Figure 12 compares the behavior achieved by every control solution. Once again, considering the existence of packet dropouts, the delay-dependent controller with no prediction stage presents the worst behavior when trying to track the nominal curves (mainly, when the curves are more pronounced). Including the prediction stage, this behavior is clearly improved, although the curves are still not accurately tracked. Using the delay-independent controller (with prediction stage), the nominal curves are precisely tracked.

# 6. - Conclusions

In this work, an NCS is presented where time-varying delays, packet dropouts and packet disorder can occur. A packet-based dual-rate control solution is proposed and defined by comparison to a previous authors' proposal based on a gain-scheduling approach [17]. Selecting the sensing period greater than the longest round-trip time delay, the packet disorder is avoided. In addition, energy saving can be achieved by reducing network load, which is a crucial issue, especially in battery-powered wireless sensors. However, in order to reach certain specifications, an $N$ times faster actuation period must be used, leading to the dual-rate control structure. Whereas the control solution in [17] (the delay-dependent controller) retunes the controller's parameters according to the round-trip time delay, the new proposal (the delay-independent controller) does not need this retuning. Both of them must include a prediction stage in order to face packet dropouts.

Simulation results reveal the superiority of the delay-independent approach, since it is able to achieve the desired (nominal) control performance. By means of a laboratory test-based platform, which uses a Cartesian robot as the process to be controlled, results are validated.

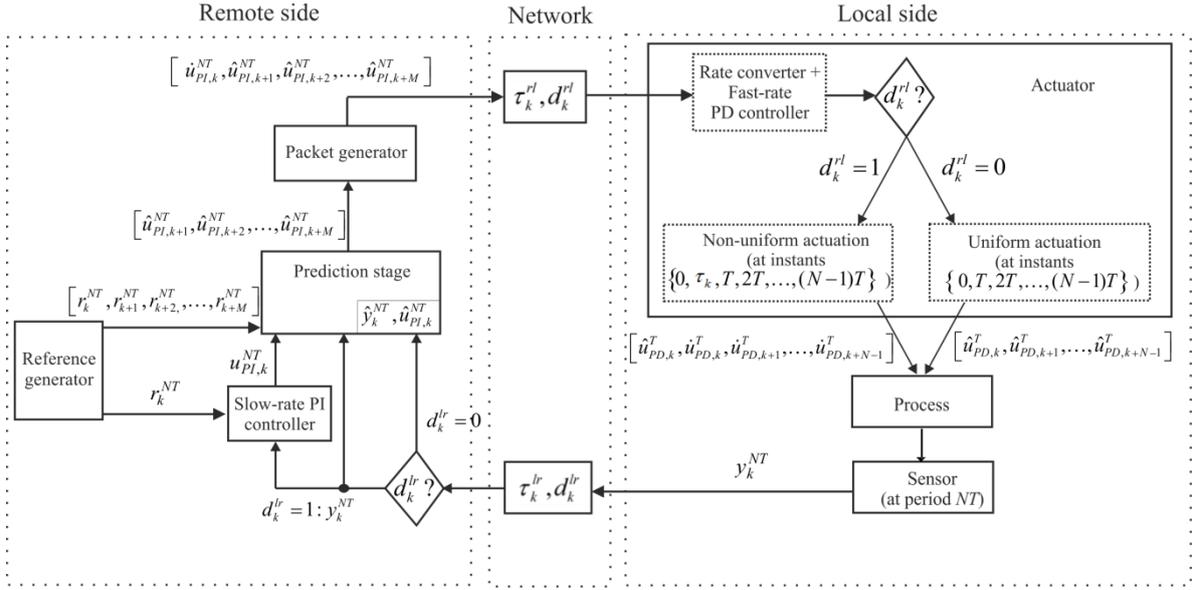

Figure 1. NCS scenario.

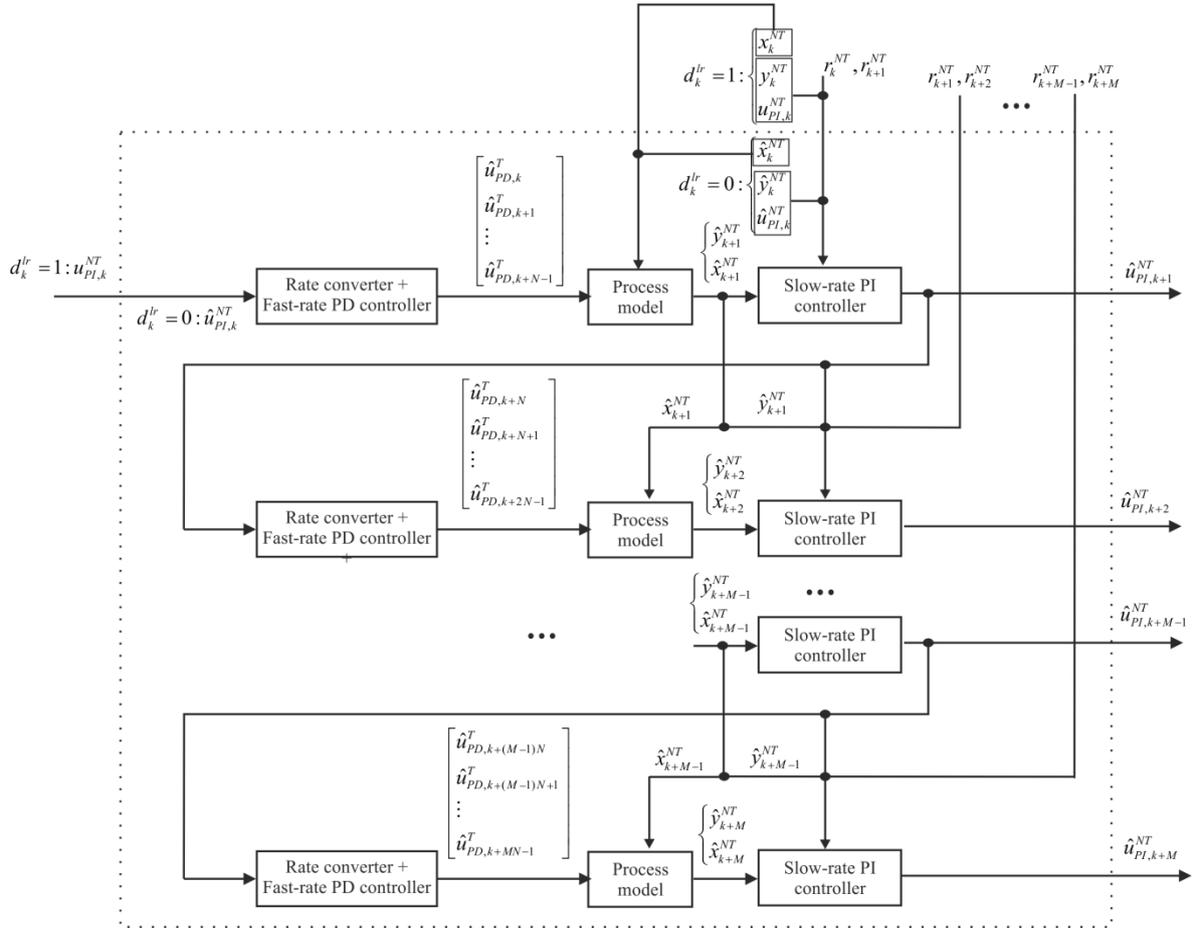

Figure 2. Prediction stage in detail.

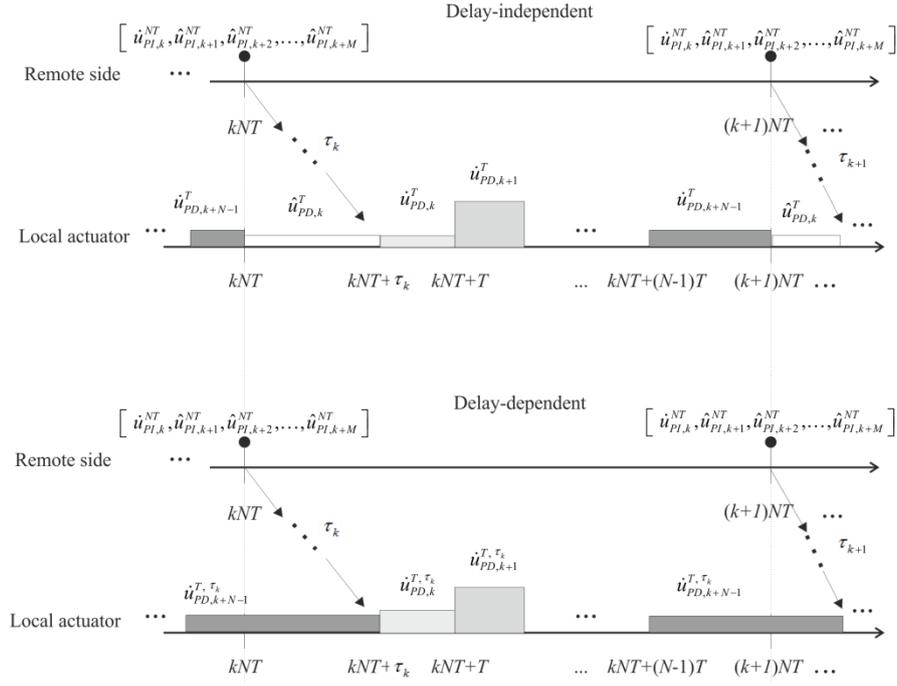

Figure 3. Comparison when no packet dropout ($d_k^{rl} = 1$).

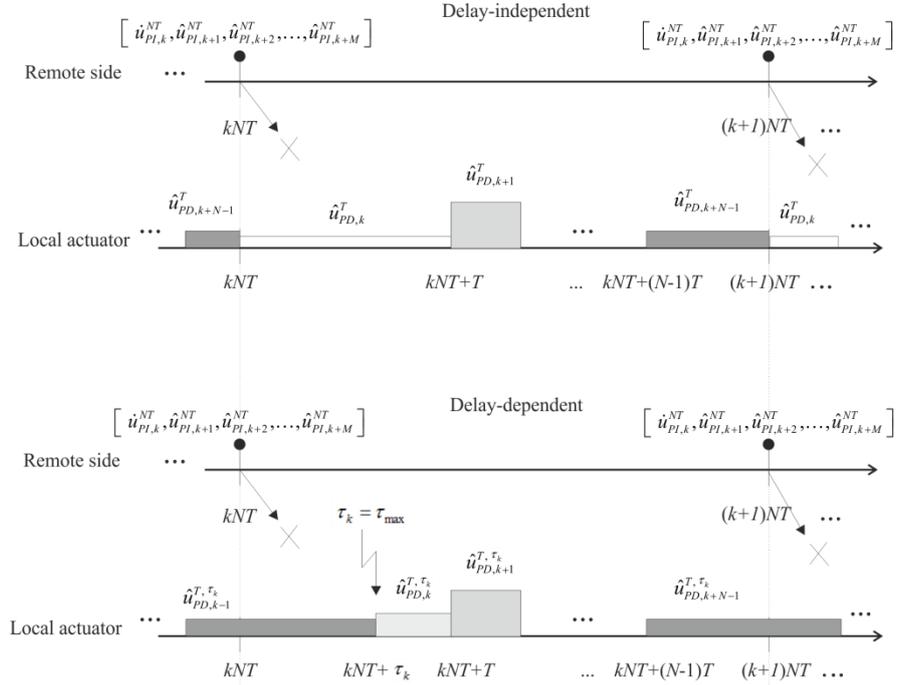

Figure 4. Comparison when packet dropout ($d_k^{rl} = 0$).

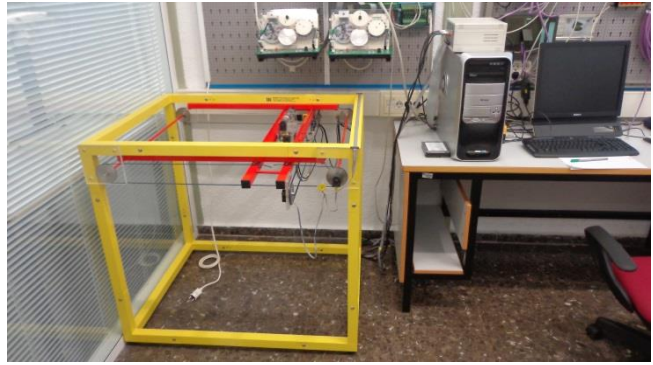

Figure 5. Cartesian robot (3D CRANE module).

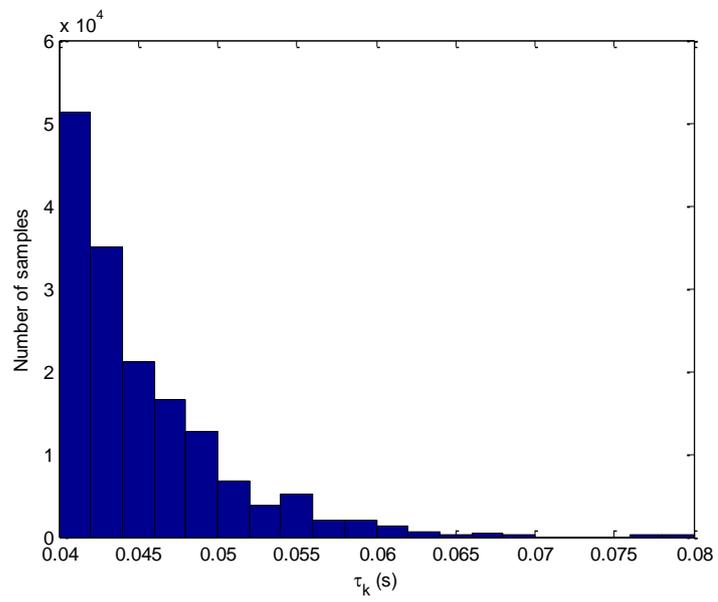

Figure 6. Delay histogram.

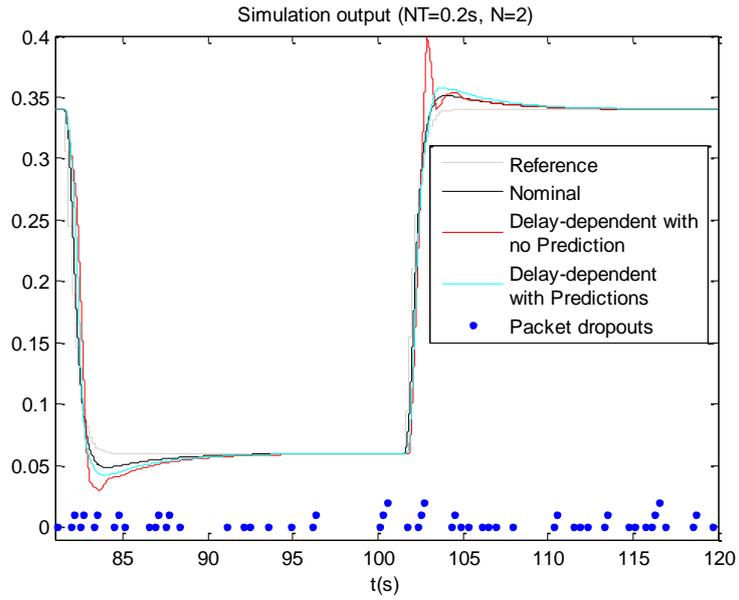

Figure 7. Comparison: nominal vs delay-dependent with packet dropouts and no prediction vs delay-dependent with packet dropouts with prediction.

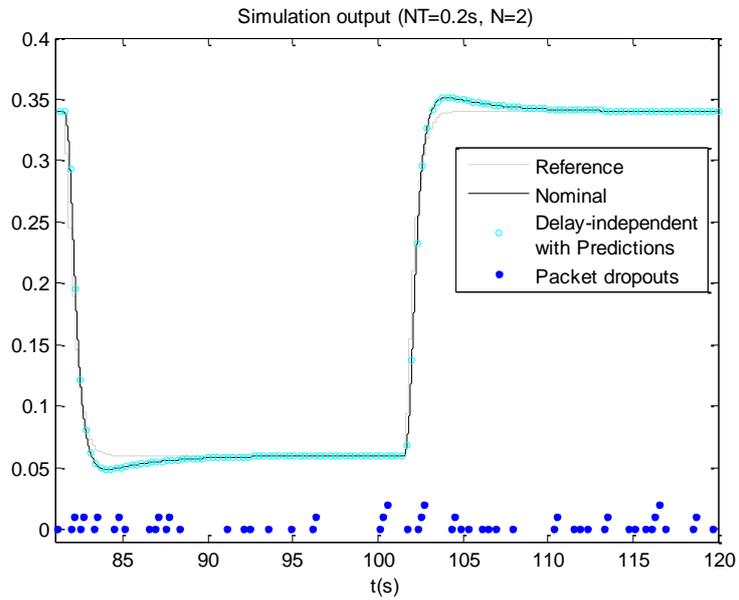

Figure 8. Comparison: nominal vs delay-independent with packet dropouts with prediction.

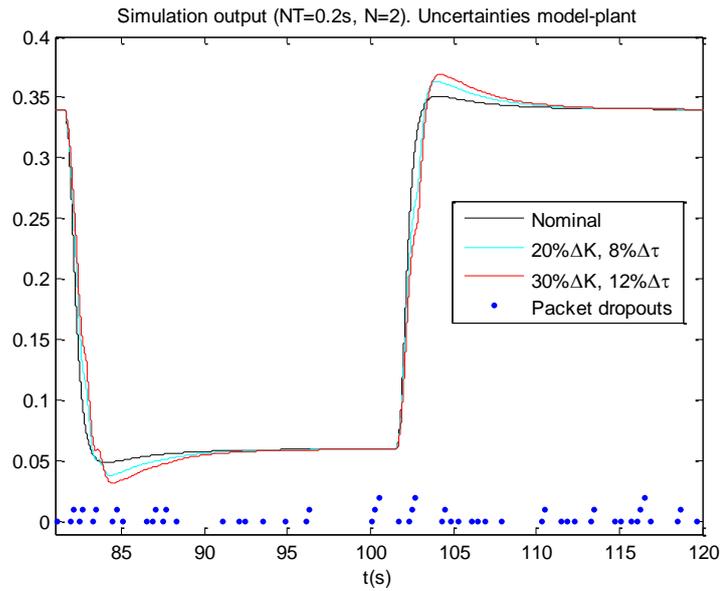

Figure 9. Comparison for uncertainties between model and plant: nominal vs delay-independent with packet dropouts with prediction.

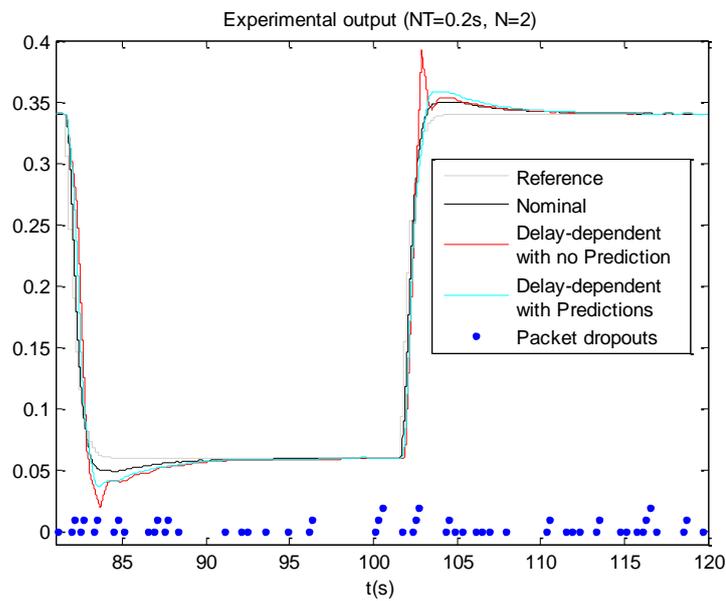

Figure 10. Comparison: nominal vs delay-dependent with packet dropouts and no prediction vs delay-dependent with packet dropouts with prediction.

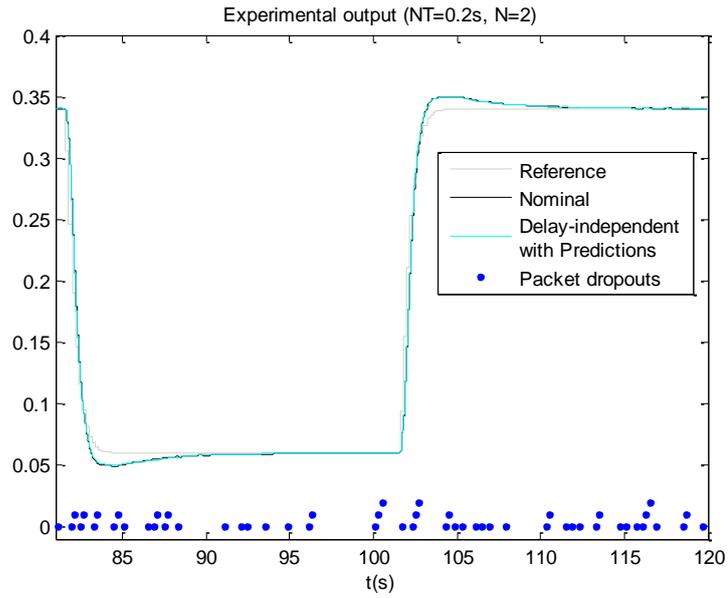

Figure 11. Comparison: nominal vs delay-independent with packet dropouts with prediction.

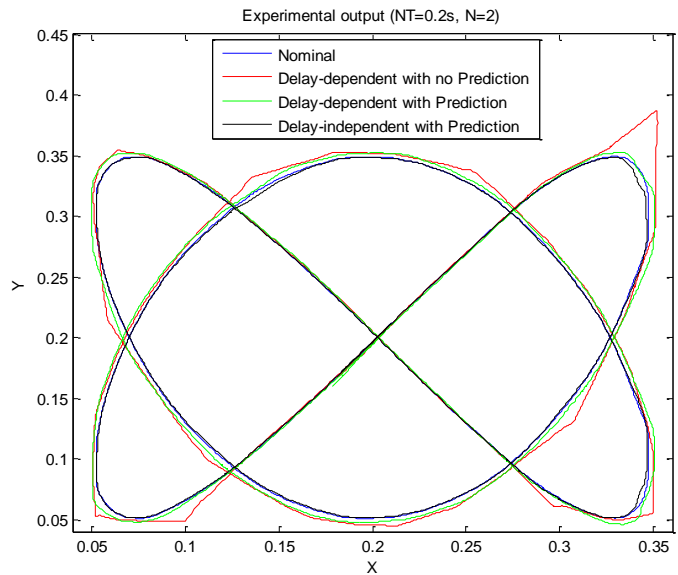

Figure 12. Lissajous curves (with packet dropouts). Comparison: nominal vs delay-dependent with no prediction vs delay-dependent with prediction vs delay-independent with prediction.

| Output | $E_Y$ | $J_1(\%)$ | $O_Y$ | $J_2(\%)$ |
|---|---|---|---|---|
| $Y_{DD-NP}$ | 291.19 | 0 | 0.047 | 0 |
| $Y_{DD-P}$ | 168.64 | 42.09 | 0.007 | 85.64 |
| $Y_{DI-P}$ | 0.02 | 99 | 0 | 100 |

Table 1. Cost indexes $J_1$ and $J_2$ in simulation.

| $E_W$ | | $q\%\Delta K$ | | | $J_3$ | | $q\%\Delta K$ | |
|---|---|---|---|---|---|---|---|---|
| | | 0 | 20 | 30 | | | 0 | 20 | 30 |
| | 0 | 0 | 183.46 | 302.60 | | 0 | 100 | 54.94 | 25.68 |
| $r\%\Delta\tau$ | 8 | 59.82 | 248.62 | 373.11 | $r\%\Delta\tau$ | 8 | 85.31 | 38.94 | 8.36 |
| | 12 | 88.56 | 280.65 | 407.19 | | 12 | 78.25 | 31.07 | 0 |

Table 2. Accumulated error $E_W$ and cost index $J_3$.

| $O_W$ | | $q\%\Delta K$ | | | $J_4$ | | $q\%\Delta K$ | |
|---|---|---|---|---|---|---|---|---|
| | | 0 | 20 | 30 | | | 0 | 20 | 30 |
| | 0 | 0 | 0.0087 | 0.0142 | | 0 | 100 | 52.19 | 21.97 |
| $r\%\Delta\tau$ | 8 | 0.0027 | 0.0117 | 0.0169 | $r\%\Delta\tau$ | 8 | 85.16 | 35.71 | 7.14 |
| | 12 | 0.0041 | 0.0130 | 0.0182 | | 12 | 77.47 | 28.57 | 0 |

Table 3. Overshoot $O_W$ and cost index $J_4$.

| Output | $E$ | $J_1(\%)$ | $O$ | $J_2(\%)$ |
|---|---|---|---|---|
| $Y_{DD-NP}$ | 290.55 | 0 | 0.042 | 0 |
| $Y_{DD-P}$ | 207.53 | 28.58 | 0.012 | 70.80 |
| $Y_{DI-P}$ | 19.53 | 93.28 | 0.001 | 97.62 |

Table 4. Cost indexes in experimentation.